\documentclass[12pt,a4paper]{article}
\usepackage{epsfig}
\begin{document}
%%%%%%%%%%%%%%%%%%%%%%%%%%%%%%%%%
\begin{titlepage}
\begin{center}
{\hbox to\hsize{\hfill UMD-PP-02-048}}

\vspace{4\baselineskip}

\textbf{\Large Neutrino oscillation data versus} \\
\bigskip
\textbf{\Large minimal supersymmetric SO(10) model }
 
\bigskip
\bigskip
\vspace{2\baselineskip}

\textbf{Takeshi Fukuyama%
\footnote{E-Mail: fukuyama@se.ritsumei.ac.jp}} \\ 
\bigskip
\textit{\small 
Department of Physics, Ritsumeikan University, Kusatsu, 
Shiga 525-8577, Japan 
}
\vspace{2\baselineskip}

\textbf{Nobuchika Okada%
\footnote{E-Mail: okadan@physics.umd.edu} }\\
\bigskip
\textit{\small
Department of Physics, University of Maryland,  
College Park, MD 20742, USA}

\vspace{3\baselineskip}

%%%%%%%%%%%%%%%%%%%%%
\textbf{Abstract}\\
\end{center}
%%%%%%%%% 
\noindent
We reconsider the minimal supersymmetric SO(10) model, 
 where only one {\bf 10} and one $\overline{\bf 126}$ 
 Higgs multiplets have Yukawa couplings 
 with matter multiplets. 
The model is generalized to include CP-violating phases, 
 and examined how well its predictions  
 can meet the current neutrino oscillation data. 
Using the electroweak scale data 
 about six quark masses, three angles and one CP-phase 
 in the Cabibbo-Kobayashi-Maskawa matrix  
 and three charged-lepton masses  
 and given $\tan \beta$ 
 (the ratio of vacuum expectation values 
  of a pair of Higgs doublets),  
 we obtain the Pontecorvo-Maki-Nakagawa-Sakata matrix 
 and the ratio, $\Delta m_\odot^2/\Delta m_\oplus^2 $, 
 as functions of only one free parameter in the model. 
In our analysis, one-loop renormalization group equations 
 for the gauge couplings, the Yukawa couplings 
 and the effective dimension-five operator 
 are used to connect the data between the electroweak scale 
 and the grand unification scale. 
Fixing the free parameter appropriately, 
 we find, for example, 
  $\sin^2 2 \theta_{12} \sim 0.72$, 
  $\sin^2 2 \theta_{23} \sim 0.90$, 
  $\sin^2 2 \theta_{13} \sim 0.16$ 
 and $\Delta m_\odot^2/\Delta m_\oplus^2 \sim 0.19$ 
 with $\tan \beta = 45$, 
 which are in agreement with 
 the current neutrino oscillation data. 
\end{titlepage}

%%%%%%%%%%%%%%%%%%%%%%%%%%%
\setcounter{footnote}{0}
%%%
\newpage
%%%%%%%%%%%%%%%%%%%%%%%%%%%

%%%%%%%%%%%%%%%%%%%%%%%
\section{Introduction} 
%%%%%%%%%%%%%%%%%%%%%%%

Supersymmetry (SUSY) extension is one of the most promising ways 
 to provide a solution to the gauge hierarchy problem 
 in the standard model \cite{review}. 
The minimal version of this extension 
 is called the minimal supersymmetric standard model (MSSM). 
Interestingly, the experimental data support 
 the unification of the three gauge couplings
 at the scale $M_{GUT} \sim 2 \times 10^{16} \mbox{GeV}$ 
 with the MSSM particle contents \cite{unification}. 
At high energies, our world may be described 
 by a SUSY grand unified theory (GUT) 
 with a simple gauge group such as SU(5) or SO(10) 
 into which all the gauge groups in the standard model 
 are embedded and unified. 

On the other hand, 
 much information about quark and lepton mass matrices 
 has been accumulated in these decades. 
Especially, there exist, at present, strong evidences 
 of neutrino masses and mixings 
 through the interpretation 
 of the (active) neutrino oscillations 
 as the solutions to the solar neutrino deficit 
 \cite{solar1, solar2, solar3, solar4, solar5, solar6}
 and the atmospheric neutrino anomaly \cite{atm}. 
According to the results, the MSSM has to be extended 
 so as to incorporate neutrino masses and mixings. 

In these points of view, 
 the supersymmetric SO(10) GUT model is 
 one of the most well-motivated theories,  
 since the right-handed neutrinos are naturally 
 incorporated and unified into the same multiplet, {\bf 16}, 
 together with the ordinary matters in the MSSM. 
Furthermore, this model has an additional advantage 
 that the smallness of the neutrino masses 
 can be naturally explained 
 through the see-saw mechanism \cite{see-saw} 
 with large right-handed Majorana neutrino masses. 

There are many possibilities 
 for the Higgs sector 
 that will make the model realistic: 
 namely, it correctly breaks the SO(10) gauge symmetry into 
 the standard model one, 
 reproduces observed fermion masses, mixings and CP-phase, 
 and realizes the doublet-triplet Higgs mass splitting  
 (in terminology of SU(5) GUT) etc. 
In this paper, we concentrate our discussion 
 on the fermion masses and mixings, 
 and consider the so-called ``minimal'' SO(10) model 
 where only one {\bf 10} and one $\overline{\bf 126}$ 
 Higgs multiplets have Yukawa couplings with 
 quarks and leptons. 
It is very interesting to investigate this model, 
 since it can fix all the fermion mass matrices 
 including, at present, ambiguous and undetermined ones. 

The minimal model was first seriously considered by 
 Babu and Mohapatra \cite{Babu-Mohapatra} (non-SUSY case) 
 and Lee and Mohapatra \cite{Lee-Mohapatra} (SUSY case), 
 where the neutrino oscillation parameters were predicted 
 with inputs of six quark masses, three mixing angles 
 in the Cabibbo-Kobayashi-Maskawa (CKM) matrix  
 and three charged-lepton masses. 
Unfortunately, the predictions were revealed 
 to contradict the current experimental data. 
For example, the predicted mixing angle, 
 $\sin^2 2 \theta_{2 3}$, 
 in the Pontecorvo-Maki-Nakagawa-Sakata (PMNS) matrix 
 was found to be too small to be consistent 
 with the the atmospheric neutrino oscillation data \cite{atm}. 
It seems to be inevitable  
 that the minimal model should be extended 
 by introducing new Higgs multiplets 
 so as to incorporate the realistic neutrino 
 oscillation parameters. 

However, note that, in Refs.~\cite{Babu-Mohapatra} 
 and \cite{Lee-Mohapatra}, 
 only the CP invariant case was analyzed. 
It was recently pointed out \cite{Matsuda-etal} that,  
 once CP-violating phases are taken 
 into account in the minimal model, 
 there exits the parameter region 
 which cannot be excluded 
 by the current neutrino oscillation data. 
This observation brings a new hope 
 to obtain the minimal SO(10) model 
 compatible with all the observed data of 
 fermion mass matrices. 
Nevertheless, more elaborate studies are needed, 
 since, in \cite{Matsuda-etal}, 
 the GUT relation among the charged fermion mass matrices 
 was applied to the mass matrices at the electroweak scale. 
This treatment is incomplete 
 because the GUT relation is valid only at the  GUT scale. 
In fact, we can see that Yukawa couplings evolve 
 according to the renormalization group equations (RGEs), 
 and the GUT relation is lost 
 after the running in the ``long desert'' 
 between the GUT scale and the electroweak scale. 

In this paper, we pursue this program 
 in more correct and complete way. 
Our strategy is the following.
First, we evaluate the data of charged fermion mass matrices 
 at the GUT scale by extrapolating the data 
 at the electroweak scale according to
 their Yukawa coupling REGs. 
Next, these GUT scale data are applied to the GUT relation 
 that is generalized to include CP-violating phases. 
This leads us to the explicit form of neutrino mass matrix 
 at the GUT scale via the GUT relation. 
Lastly, running it back to the electroweak scale 
 according to the RGE 
 for the effective dimension-five operator \cite{RGEdim5}, 
 we compare our result 
 with the current neutrino oscillation data. 

This paper is organized as follows: 
 in the next section, we give our basic framework 
 of the minimal SUSY SO(10) model 
 with one {\bf 10} and one $\overline{\bf  126}$ 
 Higgs multiplets, 
 and discuss the GUT relation among the fermion mass matrices. 
In Sec.~3, RGEs we use in our analysis are introduced. 
In Sec.~4, numerical analysis is performed, 
 and the results are presented.  
Other predictions from our resultant mass matrices 
  are discussed in Sec.~5. 
The last section is devoted to summary and comments.

%%%%%%%%%%%%%%%%%%%%%%%%%%%%%%%%%%%%%%%%%%%%%%%%%%%%%%%%%%%%%%%%
\section{Minimal SO(10) model and fermion mass matrices}
%%%%%%%%%%%%%%%%%%%%%%%%%%%%%%%%%%%%%%%%%%%%%%%%%%%%%%%%%%%%%%%%

We consider the minimal SUSY SO(10) model with a pair of 
 {\bf 10}+$\overline{\bf 126}$ Higgs multiplets, 
 only which have Yukawa couplings (superpotential) 
 with the matter multiplets such as
\begin{eqnarray}
W_Y = Y_{10}^{ij} {\bf 16}_i H_{10} {\bf 16}_j 
           +Y_{126}^{ij} {\bf 16}_i H_{126} {\bf 16}_j \; , 
\label{Yukawa1}
\end{eqnarray} 
where ${\bf 16}_i$ is the matter multiplet 
 of the i-th generation,  
 $H_{10}$ and $H_{126}$ are the Higgs multiplet 
 of {\bf 10} and $\overline{\bf 126} $ representations 
 under SO(10), respectively. 
Note that, by virtue of the gauge symmetry, 
 the Yukawa couplings, $Y_{10}$ and $Y_{126}$, 
 are complex symmetric $3 \times 3$ matrices. 
In our model, these Yukawa couplings 
 are assumed to be real as 
 in the original model \cite{Babu-Mohapatra} 
 to keep the number of free parameters minimum. 

In the SUSY GUT scenario, 
 the GUT gauge symmetry is broken at the GUT scale, 
 $M_{GUT} \sim 10^{16}$ GeV, into the standard model one. 
There are some different breaking chains 
 suitable for our aim.  
Here, suppose the intermediate pass 
 through the Pati-Salam subgroup \cite{Pati-Salam}, 
 $G_{422}=SU(4)_c \times SU(2)_L \times SU(2)_R$, 
 for simplicity. 
Under this gauge symmetry, 
 the above Higgs multiplets are decomposed as 
 ${\bf 10} \rightarrow 
 ({\bf 6},{\bf 1},{\bf 1}) + ({\bf 1},{\bf 2},{\bf 2}) $ 
 and 
 $\overline{\bf 126} \rightarrow 
 ({\bf 6}, {\bf 1}, {\bf 1} ) 
 + ( {\bf 10}, {\bf 3}, {\bf 1}) 
 + (\overline{\bf 10}, {\bf 1}, {\bf 3})  
 + ({\bf 15}, {\bf 2}, {\bf 2}) $, 
while ${\bf 16} \rightarrow ({\bf 4}, {\bf 2}, {\bf 1}) 
+ (\overline{\bf 4}, {\bf 1}, {\bf 2})$
 for the matter multiplet. 
Breaking down to the standard model gauge group, 
 $SU(4)_c \times SU(2)_R  \rightarrow SU(3)_c \times U(1)_Y$,   
 is accomplished by vacuum expectation value (VEV) 
 of the $(\overline{\bf 10}, {\bf 1}, {\bf 3})$ 
 Higgs multiplet. 
Note that the Majorana masses for the right-handed neutrinos 
 are also generated by this VEV 
 through the Yukawa coupling $Y_{126}$ 
 in Eq.~(\ref{Yukawa1}). 

After this symmetry breaking, 
 we find two pair of Higgs doublets 
 which are in the same representation as 
 the pair in the MSSM. 
One pair comes from $({\bf 1},{\bf 2},{\bf 2}) \subset {\bf 10}$ 
 and the other comes from 
 $(\overline{\bf 15}, {\bf 2}, {\bf 2}) \subset \overline{\bf 126}$. 
Using these two pairs of the Higgs doublets, 
 the Yukawa couplings of Eq.~(\ref{Yukawa1}) are rewritten as 
\begin{eqnarray}
W_Y &=& 
\overline{u}_i  \left(
 Y_{10}^{ij}  H^u_{10} + Y_{126}^{ij}  H^u_{126}     
 \right) q_j 
+
 \overline{d}_i  \left(
 Y_{10}^{ij}  H^d_{10} + Y_{126}^{ij}  H^d_{126}     
\right) q_j  \nonumber \\ 
&+&
 \overline{\nu}_i  \left(
 Y_{10}^{ij}  H^u_{10} - 3 Y_{126}^{ij} H^u_{126}   
\right) \ell_j 
+
\overline{e}_i  \left(
 Y_{10}^{ij}  H^d_{10}  - 3 Y_{126}^{ij} H^d_{126}  
\right) \ell_j   \nonumber \\
&+&
\overline{\nu}_i  
 \left( Y_{126}^{ij} \; v_R \right) 
\overline{\nu}_j \;  , 
\label{Yukawa2}
\end{eqnarray} 
where $\overline{u}$, $\overline{d}$, $\overline{\nu}$ and 
 $\overline{e}$ are the right-handed $SU(2)_L$ 
 singlet quark and lepton superfields, 
 $q$ and $\ell$ 
 are the left-handed $SU(2)_L$ doublet 
 quark and lepton superfields, 
 $H_{10}^{u,d}$ and $H_{126}^{u,d}$ 
 are up-type and down-type Higgs doublet superfields 
 originated from $H_{10}$ and $H_{126}$, respectively, 
 and the last term is the Majorana mass term 
 of the right-handed neutrinos with VEV of 
 the $(\overline{\bf 10}, {\bf 1}, {\bf 3})$ Higgs, $v_R$. 
Quarks and leptons acquire Dirac masses 
 through VEVs of these Higgs doublets. 
Note that, in that case, 
 the entry of the Clebsch-Gordan coefficient, $-3$, 
 in the lepton sector plays the crucial role  
 so that the unwanted GUT relations, 
 $m_d=m_e$ and $m_s=m_\mu$, are corrected \cite{Babu-Mohapatra} 
 in the same manner discussed by Georgi and Jarlskog 
 \cite{Georgi-Jarlskog}. 

Here, remember that the gauge coupling unification succeeds
 with only the MSSM particle contents. 
This means that only one pair of Higgs doublets remains light 
 and the others should be heavy ($\geq M_G$). 
It is highly non-trivial to accomplish this task naturally, 
 while keeping the crucial role of $H_{126}^{u,d}$ 
 to the fermion masses. 
This problem is essentially the same problem as 
 the doublet-triplet Higgs mass splitting problem, 
 and is referred  as the ``doublet-doublet splitting problem''. 
In this paper, we assume that this splitting is realized. 

Now let us see the low energy superpotential of our model. 
After realizing the doublet-doublet Higgs mass splitting, 
 we obtain only one pair of light Higgs doublets 
 ($H_u$ and $H_d$) in the MSSM, 
 which are, in general, admixture of all Higgs doublets 
 having the same quantum numbers in the original model 
 such as  
\begin{eqnarray} 
 H_u &=& \tilde{\alpha}_u  H_{10}^u  
      + \tilde{\beta}_u  H_{126}^u + \cdots 
 \nonumber \\
 H_d &=& \tilde{\alpha}_d  H_{10}^d  
      + \tilde{\beta}_d  H_{126}^d + \cdots  \; , 
 \label{mix}
\end{eqnarray} 
where ``$+ \cdots$'' denotes the mixing 
 with other Higgs doublets in the model 
 (about which we do not discuss), 
 $\tilde{\alpha}_{u,d}$ and $\tilde{\beta}_{u,d}$ 
 denote elements of the unitary matrix  
 which rotate the flavor basis in the original model 
 into the (SUSY) mass eigenstates. 
Omitting the heavy Higgs mass eigenstates, 
 the low energy superpotential is described 
 by only the light Higgs doublets $H_u$ and $H_d$ such that 
\begin{eqnarray}
W_Y &=& 
 \overline{u}_i  
 \left( \alpha^u  Y_{10}^{ij} + 
        \beta^u   Y_{126}^{ij} \right)  H_u  q_j 
+
 \overline{d}_i  
 \left( \alpha^d  Y_{10}^{ij} + 
        \beta^d   Y_{126}^{ij}  \right) H_d q_j  \nonumber \\ 
&+&
 \overline{\nu}_i  
 \left( \alpha^u  Y_{10}^{ij} -3 
        \beta^u   Y_{126}^{ij} \right)  H_u  \ell_j 
+
 \overline{e}_i  
 \left( \alpha^d  Y_{10}^{ij} -3 
        \beta^d   Y_{126}^{ij}  \right) H_d \ell_j \nonumber \\ 
 &+& \overline{\nu}_i  
   \left( Y_{126}^{ij} v_R \right)  \overline{\nu}_j \; ,  
\label{Yukawa3}
\end{eqnarray} 
where the formulas of the inverse unitary transformation 
 of Eq.~(\ref{mix}), 
 $H_{10}^{u,d} = \alpha^{u,d} H_{u,d} + \cdots $ and 
 $H_{126}^{u,d} = \beta^{u,d} H_{u,d} + \cdots $, 
 have been used. 
Note that the elements of the unitary matrix, 
 $\alpha^{u,d}$ and $\beta^{u,d}$,   
 are in general complex parameters, 
 through which CP-violating phases are introduced 
 into the fermion mass matrices. 

Providing the Higgs VEVs, 
 $H_u = v \sin \beta$ and $H_d = v \cos \beta$ 
 with $v=174 \mbox{GeV}$, 
 the quark and lepton mass matrices can be read off as 
\begin{eqnarray}
  M_u &=& c_{10} M_{10} + c_{126} M_{126}   \nonumber \\
  M_d &=&     M_{10} +     M_{126}   \nonumber \\
  M_D &=& c_{10} M_{10} -3 c_{126} M_{126}   \nonumber \\
  M_e &=&     M_{10} -3     M_{126}   \nonumber \\
  M_R &=& c_R M_{126}  \; , 
 \label{massmatrix}
\end{eqnarray} 
where the mass matrices are defined as 
 $M_{10}= Y_{10} \alpha^d v \cos\beta$ 
 and $M_{126} = Y_{126} \beta^d v \cos\beta$, respectively, 
 $c_{10}= (\alpha^u/\alpha^d) \tan \beta$, 
 $ c_{126}= (\beta^u/\beta^d) \tan \beta $, 
 and $c_R = v_R/( \beta^d  v  \cos \beta)$. 
All the quark and lepton mass matrices are characterized 
 by only two basic mass matrices, $M_{10}$ and $M_{126}$,   
 and three coefficients. 
This means the strong predictability of the model 
 to the fermion mass matrices in the MSSM. 

Now let us count how many free parameters are included 
 in the above mass matrices, not counting $c_R$. 
Remember that we began with real symmetric matrices
 $Y_{10}$ and $Y_{126}$. 
By flavor rotation of the GUT matter multiplets, 
 we can begin with the basis 
 where $M_{10}$ is real and diagonal, 
 which includes three free parameters. 
There are seven parameters in $M_{126}$: 
 six parameters originate from 
 the symmetric real matrix $Y_{126}$, 
 and one from the phase of $\beta^d$. 
Adding the four parameters from complex $c_{10}$ and $c_{126}$, 
 we find 14 free parameters in total, 
 which are evaluated as six quark masses, 
 three angles and one CP-phase in the CKM matrix
 and three charged-lepton masses, 
 but one parameter is left free \cite{Matsuda-etalI}. 
Now two free parameters are newly introduced 
 in addition to 12 free parameters 
 analyzed in \cite{Babu-Mohapatra} and \cite{Lee-Mohapatra}, 
 by which our model is generalized 
 to include CP-violating phases.

The parameter, $c_R$, appears 
 in the right-handed Majorana neutrino mass matrix, 
 by which masses of the right-handed Majorana neutrinos 
 are determined after $M_{126}$ are fixed. 
Since $v_R \sim M_G$, $c_R \sim v_R/v \sim 10^{14} \mbox{GeV}$ 
 is naturally expected. 
However, we treat $c_R$ as a free parameter 
 in the following analysis, 
 since the Higgs multiplet breaking $G_{422}$, 
 in general, mixes with other Higgs multiplets 
 being in the same representation under $G_{422}$. 
Also there should exist Higgs multiplets 
 which also have $G_{422}$ breaking VEV 
 to cancel the D-term VEV and ensure SUSY. 
In addition, threshold corrections would introduce 
 an uncertainty of $v_R$. 

The mass matrix formulas in Eq.(\ref{massmatrix}) 
 leads to the GUT relation among the quark and lepton 
 mass matrices, 
\begin{eqnarray}
M_e = c_d \left( M_d + \kappa  M_u \right) \; , 
 \label{GUTrelation} 
\end{eqnarray} 
where 
\begin{eqnarray}
 c_d &=& - \frac{3 c_{10} + c_{126}} {c_{10}-c_{126}} 
\nonumber \\
 \kappa &=& - \frac{4} {3 c_{10} + c_{126}}  \; . 
\end{eqnarray} 
Without loss of generality, we can begin with the basis 
 where $M_u$ is real and diagonal, $M_u=D_u$. 
Since $M_d$ is the symmetric matrix, 
 it can be described as $M_d = V_{KM}^* D_d V_{KM}^\dagger $ 
 by using the CKM matrix $V_{KM}$     
 and the real diagonal mass matrix $D_d$.  
Considering the basis-independent quantities, 
 $\mbox{tr}[M_e^\dagger M_e]$, 
 $\mbox{tr}[(M_e^\dagger M_e)^2]$ 
 and $\mbox{det}[M_e^\dagger M_e]$, 
 and eliminating $|c_d|$, 
 we obtain two independent equations such that 
\begin{eqnarray}
 \left(
 \frac{\mbox{tr} [ \tilde{M_e}^\dagger \tilde{M_e}] }
  {m_e^2+m_{\mu}^2+m_{\tau}^2} \right)^2
 &=& \frac{ \mbox{tr} [(
 \tilde{M_e}^\dagger \tilde{M_e} )^2 ]}
   {m_e^4+m_{\mu}^4+m_{\tau}^2} \;  ,
 \label{cond1} \\ 
 \left(
 \frac{\mbox{tr}[\tilde{M_e}^\dagger \tilde{M_e}]}
 {m_e^2+m_{\mu}^2+m_{\tau}^2} \right)^3
&=&
 \frac{\mbox{det}[\tilde{M_e}^\dagger \tilde{M_e}]}
 {m_e^2 \; m_\mu^2 \; m_\tau^2}  \; ,
 \label{cond2} 
\end{eqnarray} 
where $\tilde{M_e}= V_{KM}^* D_d V_{KM}^\dagger + \kappa D_u$. 
With the input data 
 of six quark masses, three angles 
 and one CP-phase in the CKM matrix 
 and three charged-lepton masses, 
 we can determine $\kappa$ and $|c_d|$, 
 but one parameter, the phase of $c_d$, is left undetermined 
 \cite{Matsuda-etalI}. 
Now the mass matrices, $M_{10}$ and $M_{126}$, are described by 
\begin{eqnarray}
 M_{10} &=& 
\frac{3+ |c_d| e^{i \sigma}}{4} V_{KM}^* D_d V_{KM}^\dagger
+\frac{|c_d| e^{i \sigma} \kappa}{4} D_u  \; , 
 \label{M10} \\ 
 M_{126} &=& 
\frac{1- |c_d| e^{i \sigma}}{4} V_{KM}^* D_d V_{KM}^\dagger
-\frac{|c_d| e^{i \sigma} \kappa}{4} D_u \; , 
 \label{M126} 
\end{eqnarray} 
 as the functions of $\sigma$, the phase of $c_d$, 
 with the solutions, $|c_d|$ and $\kappa$, 
 determined by the GUT relation. 

Note that the GUT relation of Eq.~(\ref{GUTrelation})  
 is valid only at the GUT scale. 
Therefore, the input data 
 needed to determine $\kappa$ and $|c_d|$  
 should be the data at the GUT scale. 
To evaluate the GUT scale data, 
 RGE analysis of Yukawa couplings is necessary. 
This is the subject of the next section.

%%%%%%%%%%%%%%%%%%%%%%%
\section{One-loop RGEs} 
%%%%%%%%%%%%%%%%%%%%%%%

The data we need at the GUT scale 
 are six quark masses, three angles 
 and one CP-phase in the CKM matrix 
 and three charged-lepton masses. 
In general, it is not a straightforward task 
 to obtain the data at GUT scale 
 by extrapolating the data at the electroweak scale. 
This is because RGEs of quark and charged-lepton 
 Yukawa couplings include unknown neutrino Yukawa couplings. 
We can obtain a reliable answer 
 only if neutrino Yukawa couplings are negligible. 

Fortunately, in models with the see-saw mechanism, 
 the right-handed neutrinos are decoupled at the scale 
 below the heavy Majorana masses, and  
 the neutrino Yukawa couplings become 
 irrelevant and disappear from RGEs 
 of quark and charged-lepton Yukawa couplings. 
This fact makes our analysis straightforward, 
 at least, below the heavy Majorana mass scale. 
In the following analysis, we use the RGEs 
 independent of the neutrino Yukawa couplings 
 at all scales between the electroweak and the GUT scales. 
Although this treatment fails at the scale 
 above the right-handed Majorana neutrino masses, 
 we expect that our results are reliable,  
 since the heavy Majorana mass scale 
 is close to the GUT scale. 
We will give a comment on this point in the last section. 
 
We analyze one-loop RGEs for gauge couplings 
 and quark and charged-lepton Yukawa couplings in the MSSM. 
For simplicity, the universal soft SUSY breaking masses 
 are assumed, $m_{SUSY} = M_Z \sim 91 \mbox{GeV}$. 
In this case, the GUT scale is found to be 
 $M_G \sim 2 \times 10^{16} \mbox{GeV}$. 
The RGEs used in our analysis are as follows \cite{RGEY}. 
For gauge couplings, 
\begin{eqnarray}
\frac{d g_i}{d (\ln \mu)} = - \frac{b_i}{16 \pi^2} g_i^3  \; , 
\end{eqnarray} 
where $b_i=(3, -1, -33/5)$ for $i=3,2,1$ 
 corresponding to $SU(3)_c$, $SU_L$ and $U(1)_Y$ 
 gauge couplings, respectively. 
We use the electroweak scale inputs 
 such that $\alpha_3(M_Z) = g_3^2 / 4 \pi= 0.118$,  
 $\alpha_2(M_Z) = g_3^2 / 4 \pi= 0.0335$ 
 and $\alpha_1(M_Z) = g_3^2 / 4 \pi= 0.0168$.  
On the other hand, RGEs for the Yukawa couplings of 
 up-type quarks ($Y_u$), down-type quarks ($Y_d$), 
 and charged-leptons ($Y_e$) are found to be 
\begin{eqnarray}
\frac{d S_F}{d (\ln \mu)} 
=\frac{1}{16 \pi^2} \left( \beta_F S_F + S_F \beta_F 
\right) \; ,  
\end{eqnarray}   
where $S_F=Y_F^\dagger Y_F$ with fermion index $F=u,d,e$, 
and 
\begin{eqnarray}
 \beta_u &=& 3 S_u +S_d 
+ \left\{  \mbox{tr}\left[3  S_u \right] 
- \left( \frac{16}{3} g_3^2 + 3 g_2^2 + \frac{13}{15} g_1^2
\right) \right\} {\bf 1}_{3 \times 3}   \nonumber \\ 
 \beta_d &=& 3 S_d +S_u  
+ \left\{  \mbox{tr}\left[ 3 S_d + S_e \right] 
- \left(\frac{16}{3} g_3^2 + 3 g_2^2 + \frac{7}{15} g_1^2
\right) \right\} {\bf 1}_{3 \times 3}   \nonumber \\ 
 \beta_e &=& 3 S_e 
+ \left\{ \mbox{tr} \left[3  S_d + S_e \right] 
- \left(3 g_2^2 + \frac{9}{5} g_1^2
\right) \right\} {\bf 1}_{3 \times 3}   \; .     
 \label{RGEY}
\end{eqnarray} 
To determine the input values of fermion Yukawa couplings 
 at $M_Z$, we have to fix $\tan \beta$ by hand. 
For given $\tan \beta$, 
  they are given by 
\begin{eqnarray}
S_u &=& \mbox{diag}\left( 
\frac{m_u^2}{v^2}, \frac{m_c^2}{v^2}, 
\frac{m_t^2}{v^2}  \right)  
\left( 1+ \tan^{-2} \beta \right) \; , \\
S_d &=& V_{KM} \;  \mbox{diag}\left( 
 \frac{m_d^2}{v^2}, \frac{m_s^2}{v^2}, 
 \frac{m_b^2}{v^2}  \right)  V_{KM}^\dagger 
 \left( 1+ \tan^2 \beta \right) \; , \\ 
S_e &=& \mbox{diag}\left( 
\frac{m_e^2}{v^2}, \frac{m_\mu^2}{v^2}, 
\frac{m_\tau^2}{v^2}  \right)  
\left( 1+ \tan^{-1}\beta \right) \; .  
\end{eqnarray} 
Here, we have worked out in the basis 
where $S_u$ and $S_e$ are diagonal. 
Given $\tan \beta$, it is easy 
 to solve the above RGEs numerically. 
Although, as can be seen from the above RGEs, 
  $S_u$ is no longer diagonal after running, 
  we can extract the data we need at the GUT scale 
  by the usual manner. 
For example, the CKM matrix at the GUT scale is 
 given by $V_{KM} = V_u^\dagger V_d$, 
 where $V_u$ and $V_d$ are the unitary matrices 
 which diagonalize $S_u$ and $S_d$, respectively. 

Solving the GUT relation as discussed in the previous section 
 by using the extrapolated data, 
 we can obtain the light Majorana neutrino mass matrix, 
 $M_{\nu}$, as a function of $\sigma$ and $c_R$ 
 through Eqs.~(\ref{massmatrix}), (\ref{M10}) and (\ref{M126}) 
 and the see-saw mechanism, $M_{\nu}=-M_D^T M_R^{-1} M_D$. 
Since this is the mass matrix at the GUT scale, 
 it should be extrapolated to the electroweak scale 
 according to the RGE 
 for the effective dimension-five operator
 \cite{RGEdim5} such that 
\begin{eqnarray}
\frac{d M_{\nu}}{d (\ln \mu)} 
= \frac{1}{16 \pi^2} 
\left\{ 
S_e^T M_{\nu} + M_{\nu} S_e 
 + \left(6 \mbox{tr}\left[ S_u \right]- 6 g_2^2-
  \frac{6}{5} g_1^2
   \right) M_{\nu} \right\} \; . 
\label{RGEN}
\end{eqnarray} 
After the running, our resultant neutrino mass matrix 
 are compared with the neutrino oscillation data. 
In the next section, 
 the effect of the RGE running will be found 
 to be small and almost negligible in our model.

%%%%%%%%%%%%%%%%%%%%%%%%%%%%%%%%%%%%%%%%%%
\section{Numerical analysis and results} 
%%%%%%%%%%%%%%%%%%%%%%%%%%%%%%%%%%%%%%%%%%

Now we are ready to perform all the numerical analysis. 
Fist of all, note that the solution of the GUT relation exists, 
 only if we take appropriate input parameters. 
This is because the number of our free parameters (fourteen) 
 are almost the same as the number of inputs (thirteen). 
Therefore, in the following analysis, 
 we vary two input parameters, 
 $m_s$ and CP-phase $\delta$ in the CKM matrix, 
 within the experimental errors so as to find the solution. 
We take input fermion masses at $M_Z$ as follows (in GeV): 
\begin{eqnarray}
& & m_u = 0.00233 \; , \; \;   m_c=0.677 \; , 
\; \;  m_t=176 \nonumber \\
& & m_d=0.00469 \; ,  \; \;  m_b=3.00  \nonumber \\ 
& & m_e=0.000487 \; , \; \; m_\mu=0.103 \; ,
\; \;  m_\tau=1.75 \nonumber 
\end{eqnarray} 
Here, the experimental values extrapolated from low energies 
 to $M_Z$ were used \cite{Fusaoka-Koide}. 
For the CKM mixing angles in the ``standard'' parameterization, 
 we input the center values measured by experiments 
 as follows \cite{PDG}: 
\begin{eqnarray}
 s_{12} = \frac{0.219+0.226}{2} , \; \;  
 s_{23} = \frac{0.037+0.043}{2} , \; \;
 s_{13} = \frac{0.002+0.005}{2} \; . \nonumber 
\end{eqnarray} 

Since it is very difficult to search all the possible
 parameter region systematically, 
 we present only reasonable results we found. 
In the following, we show our analysis in detail 
 in the case $\tan \beta=45$. 
Here we take $m_s= 0.0729 \mbox{(GeV)}$ 
 and $\delta= 86.4^\circ$. 
In this case, the CKM matrix at $M_Z$ is given by 
\begin{eqnarray}
 V_{KM}(M_Z) = \left( 
 \begin{array}{ccc}
0.975 & 0.222 & 0.000220-0.00349i \\
-0.222-0.000136 i &  0.974-0.0000311 i & 0.0400 \\ 
0.00869-0.00340 i & -0.0390-0.000777 i & 0.999
   \end{array} 
  \right) \; .  \nonumber 
\end{eqnarray}
After RGE evolutions to the GUT scale, 
 fermion masses (up to sign) and the CKM matrix 
 are found as follows (in GeV) : 
\begin{eqnarray}
& & m_u = 0.00103 \; , \; \;   m_c=0.301 \; , 
\; \;  m_t=134 \nonumber \\
& & m_d=0.00171 \; , \; \;  m_s =0.0265   \; , 
\; \;   m_b=1.56  \nonumber \\ 
& & m_e=0.000413 \; , \; \; m_\mu=0.0872 \; ,
\; \;  m_\tau=1.69  \nonumber 
\end{eqnarray} 
and 
\begin{eqnarray}
 V_{KM}(M_G) = \left( 
 \begin{array}{ccc}
0.975 & 0.222 & 0.000175 - 0.00279 i \\
-0.222-0.000121 i & 0.974 + 0.000129 i & 0.0320 \\ 
0.00695 - 0.00272 i & -0.0312 - 0.000626 i & 0.999
   \end{array} 
  \right) \;   \nonumber
\end{eqnarray}
 in the standard parameterization. 

These outputs at the GUT scale are used 
 as input parameters 
 in order to solve Eqs.~(\ref{cond1}) and (\ref{cond2}).  
The contours of solutions of each equations 
 are depicted in Fig.~1, 
 where the signs of the input fermion masses have taken as 
 $-$ for $m_u$, $m_c$, $m_d$, and $m_s$, 
 and $+$ for $m_t$ and $m_b$. 
We find a solution 
\begin{eqnarray}
 & \kappa =
-0.0134 + 0.000786 i 
\nonumber \; , & \\
 & |c_d| = 6.38  \; , & 
\end{eqnarray}
 which leads to the reasonable results 
 for the neutrino sector.% 
\footnote{
Although there are other solutions, we do not address them. 
They lead to small $\sin^2 2 \theta_{2 3}$ 
 inconsistent with the atmospheric neutrino oscillation data,
 and are out of our interest. }

Now we can describe all the fermion mass matrices 
 as functions of $\sigma$ and $c_R$ 
 by using the above solution 
 through the mass matrix formulas 
 of Eqs.~(\ref{massmatrix}), (\ref{M10}) and (\ref{M126}). 
Since the PMNS mixing angles are independent of $c_R$, 
 we can plot the angles as functions of $\sigma$. 
The results are depicted in Fig.~2, 
 and we can see that the resultant mixing angles are 
 very sensitive to $\sigma$. 
For $\sigma=3.198 [\mbox{rad}] $, we obtain   
 $\sin^2 2 \theta_{1 2} = 0.722$, 
 $\sin^2 2 \theta_{2 3} = 0.881$ 
 and $\sin^2 2 \theta_{1 3} = 0.164$ at the GUT scale. 
After running this results back to the electroweak scale 
 according to RGE of Eq.~(\ref{RGEN}), we find 
\begin{eqnarray}
 \sin^2 2 \theta_{1 2} = 0.723, \; \;  
 \sin^2 2 \theta_{2 3} = 0.895, \; \;  
 \sin^2 2 \theta_{1 3} = 0.164 \; .   
\end{eqnarray}  
Note that RGE running effects are almost negligible. 
The ratio $ \Delta m_{\odot}^2/ \Delta m_{\oplus}^2 $ 
 is also independent of $c_R$, 
 where $\Delta m_{\odot}^2$ and $ \Delta m_{\oplus}^2$ 
 are the oscillation parameters relevant for 
 the solar and the atmospheric neutrino deficits, respectively. 
In Fig.~3, the ratio at the GUT scale is depicted 
 as a function of $\sigma$. 
After the RGE running, we find, at the electroweak scale,  
\begin{eqnarray} 
 \frac{\Delta m_\odot^2}{\Delta m_\oplus^2} = 0.188 \; . 
\end{eqnarray}  
RGE running effects are almost negligible 
 also for the ratio. 
The neutrino mass matrix at the electroweak scale 
 and the PMNS matrix 
 which lead to the above results are as follows: 
\begin{eqnarray}
 M_{\nu} = c_R^{-1} 
\left( 
 \begin{array}{ccc}
14.7+48.0 i & -46.9-96.5 i & -446-370 i \\
-46.9-96.5 i & -347+90.3 i & 1030+702 i \\
-446-370 i & 1030+702 i & -497-914 i 
  \end{array} 
  \right) \; (\mbox{GeV}) \; ,  \nonumber
\end{eqnarray} 
and 
\begin{eqnarray}
 U_{PMNS} = \left( 
 \begin{array}{ccc}
0.168 + 0.838 i & -0.467+0.0940 i & -0.00508+0.207 i \\
0.0519 + 0.498 i & 0.651-0.0473 i & 0.0189 -0.569 i \\ 
0.0745 + 0.116 i & 0.450-0.381 i & 0.431 + 0.669 i 
   \end{array} 
  \right) \; .  \nonumber
\end{eqnarray}
They have been calculated in the basis 
 where the charged-lepton mass matrix is diagonal 
 and all the matrix elements are real and positive. 
The neutrino mass eigenvalues are determined 
 by fixing $c_R$ appropriately. 
For example, taking $c_R=3.01 \times 10^{13}$ 
 leads to $\Delta m_\odot^2=3.76 \times 10^{-4} \mbox{eV}^2$ 
 and $\Delta m_\oplus^2=2.00 \times  10^{-3} \mbox{eV}^2$. 
In this case, three mass eigenvalues of 
 the right-handed neutrinos are found to be (in GeV)  
 $M_{R_1}=1.64 \times 10^{11} $, 
 $M_{R_2}=2.50 \times 10^{12} $ and  
 $M_{R_3}=8.22 \times 10^{12} $, respectively. 
These results are in agreement with 
 the results of the recent neutrino oscillation analysis 
 \cite{recent-analysis} with the large angle solution 
 with Mikheev-Smirnov-Wolfenstein effect \cite{MSW}
 for the solar neutrino deficit. 
 
Our results are very sensitive to the input values 
 of $m_s$ and $\delta$. 
As mentioned above, the solution 
 of Eqs.~(\ref{cond1}) and (\ref{cond2}) exists 
 only if the input values of 
 $m_s$ and $\delta$ are appropriately fixed. 
In Fig.~1, as $m_s$ is taken to be larger for fixed $\delta$, 
 the contour of Eq.(\ref{cond1}) moves to the left, 
 while the circle moves to the right. 
Thus, the solution disappears eventually. 
 On the other hand, 
 as $\delta$ is taken to be larger for fixed $m_s$, 
 both of the contours move to the left, 
 but the circle moves faster. 
Eventually, the solution disappears. 
This disappearance of the solution occurs 
 by several percent changes of the input parameters. 
Therefore, with fixed $m_s$ ($\delta$), 
 $\delta$ ($m_s$) is almost determined 
 in order for the solution to exist. 

Next let us see a more important parameter dependence. 
When $\delta$ and $m_s$ are fixed appropriately 
 and the solutions are found, 
 we obtain the similar plot as in Fig.~2 
 for the mixing angles in the PMNS matrix. 
However, the hight of the peaks depends on 
 the input $\delta$ values. 
As in the past works \cite{Babu-Mohapatra} \cite{Lee-Mohapatra}, 
 we mostly find that the resultant $\sin^2 2 \theta_{2 3}$ 
 is too small to be consistent with the experimental data. 
Only the special values of the inputs and $\sigma$  
 can give the results being in agreement 
 with all the neutrino oscillation data. 
The reasonable results we found are as follows: 
\begin{center}
\begin{tabular}{c|cc|c|ccc|c}
 $\tan \beta $ & $m_s(M_Z)$ & $\delta$  & $\sigma $
 & $\sin^2 2 \theta_{1 2}$
 & $\sin^2 2 \theta_{2 3}$
 & $\sin^2 2 \theta_{1 3} $
 & $\Delta m_{\odot}^2/\Delta m_{\oplus}^2$ \\ \hline
40 & 0.0718 & $ 93.6^\circ $ & 3.190& 
0.738 & 0.900 & 0.163 & 0.205 \\
45 & 0.0729 & $ 86.4^\circ $ & 3.198& 
0.723 & 0.895 & 0.164 & 0.188 \\
50 & 0.0747 & $ 77.4^\circ $ & 3.200& 
0.683 & 0.901 & 0.164 & 0.200 \\
55 & 0.0800 & $ 57.6^\circ $ & 3.201& 
0.638 & 0.878 & 0.152 & 0.198 \\
\end{tabular}
\end{center}
The mass eigenvalues of the light Majorana neutrinos 
 are determined by $c_R$. 
In other words, $c_R$ is a function of 
 the oscillation parameter, 
 and appropriate value of $c_R$ is fixed 
 so as to be suitable for the neutrino oscillation data. 
These relations are listed in the following. 
\begin{center}
\begin{tabular}{c|cc}
 $\tan \beta $ & 
 $c_R \times m_{\nu_3} \mbox{(TeV)}~$ &
 $ ~c_R \times 
 \sqrt{\frac{\Delta m_\oplus^2}{10^{-3} \mbox{eV}^2}} $ 
 \\   \hline
40 & 2.18~ & $~6.26 \times 10^{13}$  \\
45 & 2.08~ & $~6.02 \times 10^{13}$  \\
50 & 1.81~ & $~5.23 \times 10^{13}$  \\
55 & 1.42~ & $~4.04 \times 10^{13}$  \\
\end{tabular}
\end{center}
Here, $m_{\nu_3}$ is the mass eigenvalue of the heaviest 
 light Majorana neutrino. 
As mentioned above, 
 our resultant neutrino oscillation parameters 
 are sensitive to all the input parameters. 
In other words, if we use the neutrino oscillation data 
 as the input parameters, 
 the other input, for example, the CP-phase in the CKM matrix 
 can be regarded as the prediction of our model. 
It is a very interesting observation 
 that the CP-phases listed above are 
 in the region consistent with experiments \cite{PDG}.

%%%%%%%%%%%%%%%%%%%%%%%%%%%%%%%%%%
\section{Other predictions} 
%%%%%%%%%%%%%%%%%%%%%%%%%%%%%%%%%%

Now all the mass matrices have been completely determined 
 in our model as discussed in the previous section. 
There are other interesting physical observables 
 which can be calculated 
 by using the concrete fermion mass matrices in our model. 

The averaged neutrino mass relevant 
 for the neutrino-less double beta decay \cite{Doi-etal}
 can be read off from  $(1,1)$ component of
 the light Majorana neutrino mass matrix $M_\nu$.  
The CP-violation in the lepton sector
 is characterized by the Jarlskog parameter \cite{Jarlskog} 
 defined as 
\begin{eqnarray}
 J_{CP} = \mbox{Im}\left[ 
     U_{e2} U_{\mu 2}^*  U_{e 3}^*  U_{\mu 3} \right] \; , 
\end{eqnarray}  
where $U_{f i}$ is the PMNS matrix element. 

It is well known that the SO(10) GUT model possesses 
 a simple mechanism of baryogenesis 
 through the out-of-equilibrium decay 
 of the right-handed neutrinos, 
 namely, the leptogenesis \cite{Fukugita-Yanagida}. 
The amount of the created baryon asymmetry 
 is characterized by the CP-violating parameter $\epsilon$ 
 estimated as 
\begin{eqnarray}  
  \epsilon \sim 
 \sum_{j=2,3} \frac{\mbox{Im} \left[  
 (M_D M_D^\dagger)_{1j}^2  \right]}
 {v^2 \sin^2 \beta  (M_D M_D^\dagger)_{11} } 
 \frac{M_{R_1}}{M_{R_j}}  \; . 
\end{eqnarray}  
Here, $M_D$ is the neutrino Dirac mass matrix 
 in the basis where the right-handed Majorana mass matrix 
 $M_R$ is real and diagonal, 
 and $M_{R_j}$ is the mass eigenvalue of 
 the right-handed Majorana neutrino of the i-th generation. 

These quantities are evaluated by using the results 
 presented in the previous section, 
 and results are listed in the following.  
\begin{center}
\begin{tabular}{c|ccc}
 $\tan \beta $ & 
 $ \langle m_\nu \rangle_{ee}~ (\mbox{eV})$ & 
 $J_{CP}$ &   $\epsilon$  \\   \hline
40 & 0.00122  & $~~0.00110$ & $ 7.39 \times 10^{-5} $ \\
45 & 0.00118  & $-0.00429$  & $ 6.80 \times 10^{-5} $ \\
50 & 0.00119  & $-0.00631$  & $ 6.50 \times 10^{-5} $ \\
55 & 0.00117  & $-0.00612$  & $ 11.2 \times 10^{-5} $ \\ 
\end{tabular}
\end{center}
Here, $c_R$ was fixed so that 
 $\Delta m_\oplus^2 = 2 \times 10^{-3} \mbox{eV}^2$. 
Unfortunately, both of the averaged neutrino mass 
 and the Jarlskog parameter 
 may be too small to expect their evidences  
 in future experiments. 
On the other hand, the CP-violating parameter 
 is too large to be consistent with 
 the observed baryon asymmetry. 
The problem of this too large baryon asymmetry 
 can be easily avoided by assuming an inflationary universe 
 whose reheating temperature is smaller than 
 the right-handed neutrino masses. 
However, in this case, the leptogenesis scenario 
 no longer works, and another scenario of baryogenesis 
 such as the Affleck-Dine mechanism \cite{Affleck-Dine} 
 may be applicable. 

Sizable lepton-flavor violation (LFV) 
 can be expected in SUSY models, and 
 the LFV processes are one of the most important 
 processes as the low-energy SUSY search \cite{Hisano}. 
In detailed analysis, 
 concrete fermion mass matrices are necessary. 
It is worth investigating the LFV processes in our model.

%%%%%%%%%%%%%%%%%%%%%%%%%%%%%%%%%%
\section{Summary and comments} 
%%%%%%%%%%%%%%%%%%%%%%%%%%%%%%%%%%

We have discussed the minimal SUSY SO(10) model, 
 where only one {\bf 10} and one $\overline{\bf 126}$ 
 Higgs multiplets have Yukawa couplings with fermions. 
This model can determine all the fermion mass matrices 
 with only a few free parameters. 
It is known that, in the absence of CP violation, 
 this model cannot incorporate 
 the realistic neutrino mass matrix 
 consistent with the neutrino oscillation data.  
We examined more general case in which CP-violating phases 
 in the fermion mass matrices were introduced. 
In this case, using the experimental data 
 of six quark masses, three angles and one CP-phase 
 in the CKM matrix and three charged-lepton masses, 
 the light Majorana neutrino mass matrix can be determined  
 as the function of only one free parameter $\sigma$ 
 through the GUT relation and the see-saw mechanism. 
Here we do not count $c_R$, which plays only the role 
 to fix the overall scale of the neutrino masses. 
In order to connect the mass matrix data 
 between the electroweak and the GUT scales, we have analyzed 
 the one-loop RGEs of the charged fermion Yukawa couplings 
 and the effective dimension-five operator 
 of the light Majorana neutrinos. 

We found that there was the parameter region  in which 
 the predicted neutrino mass matrix can be consistent 
 with the current neutrino oscillation data. 
This parameter region is severely constrained, 
 and all the parameters, even the CP-phase in the CKM matrix,  
 are almost fixed for given $\tan \beta$. 
In other words, we can regard the neutrino oscillation data 
 as the inputs while the CP-phase as the output. 
Interestingly, our results consistent 
 with the neutrino oscillation data 
 are also consistent with experimental results 
 of CP-violation in the quark sector.  

In our RGE analysis, we have used RGEs introduced in Sec.~3 
 at all the energy scales between the electroweak scale  
 and the GUT scale. 
This treatment fails at the energy larger 
 than the right-handed Majorana neutrino mass scale, 
 where the right-handed neutrinos are no longer decoupled. 
Above that scale, we have to take RGE of 
 neutrino Yukawa coupling into account. 
However, as found in Sec.~4, the Majorana mass scale 
 is about $10^{13} \mbox{GeV}$, 
 and close to the GUT scale 
 in the sense of the RGE (logarithmic) running.  
Thus, we expect that our results can be still reliable. 
In fact, we can find that changes of our input values 
 at the GUT scale remain within several percents 
 even if the neutrino Yukawa coupling RGE 
 is taken into account, 
 and that the results presented in Sec.~4 
 are almost unchanged.  

In addition, we have neglected the SUSY threshold corrections 
 in our RGE analysis. 
It is known that with large $tan \beta$ 
 the down-type quark mass matrix is potentially affected 
 by large SUSY threshold corrections 
 and we can neglect them with a limited region 
 of the soft SUSY breaking parameters \cite{SUSY-threshold}. 
If the SUSY threshold corrections are large, 
 it should be taken into account in our analysis, 
 and will change the input values at the GUT scale 
 and eventually the final result. 
It is an interesting question 
 whether the result gives a better or worse fit,   
 and it is worth analyzing the RGEs  
 taking SUSY threshold  corrections into account.  

When the CHOOZ data \cite{CHOOZ} is concerned, 
 our results seems to be in sever situation 
 \cite{Bilenky-etal}. 
However, more comprehensive analysis \cite{Gonzalez-Garcia} 
 gives rather loose constraint. 
In any case, our results lie in the allowed region at 99\% C.L.  
 in the global analysis \cite{recent-analysis}, 
 and we have concluded that they are in agreement 
 with the current neutrino oscillation data.%
\footnote{
We thank Osamu Yasuda for his analysis concluding that 
 our results is 2.6 $\sigma$ away from the current best fit value. 
}
This situation seems to be inevitable in our model, 
 since two angles, $\sin^2 2 \theta_{23}$ and 
 $\sin^2 2 \theta_{13}$, are strongly correlated 
 and have the peaks at almost the same $\sigma$ value 
 as can be seen in Fig.~2. 
This can be easily avoided, 
 when we extend the minimal model and introduce new Higgs multiplets. 
An extended model has new free parameters in fermion mass matrices, 
 and can be in much better agreement with the neutrino oscillation data, 
 larger $\sin^2 2 \theta_{23}$ but smaller $\sin^2 2 \theta_{13}$. 
However, what we have found in this paper is that 
 the minimal SO(10) model, the model with only minimal set of Higgs 
 multiplets, is still viable without any extention.

We have concentrated our discussion only on the Yukawa sector 
with one ${\bf 10}$ and one $\overline{\bf 126}$ Higgs multiplets. 
In order to complete our model, it is necessary to construct 
a concrete Higgs sector which can realize all the assumptions 
in our model, the correct GUT symmetry breaking, 
doublet-doublet (triplet) Higgs mass splitting etc.  
Such a Higgs sector may include new Higgs multiplets 
which affect the fermion mass matrices. 
If it is the case, our model becomes less predictive, 
although the model can be in better agreement with 
the neutrino oscillation data.% 
\footnote{
In this point of view, the Higgs sector proposed in \cite{Lee-Mohapatra} 
is interesting, because it remains the Yukawa sector minimal.}  
This issue is highly non-trivial, and a further work is needed. 

Although, in this paper, the SUSY SO(10) model has been considered, 
 non-SUSY SO(10) model with intermediate symmetry breaking scales 
 is also worth investigating.  
In this case, the strategy is completely analogous 
 to that we discussed in this paper, 
 except that RGEs are replaced by the ones of non-SUSY case. 

%%%%%%%%%%%%%%%%%%%%%%%%%%%%%%%
\section*{Acknowledgments}
%%%%%%%%%%%%%%%%%%%%%%%%%%%%%%%
We would like to thank Rabindra N. Mohapatra  
 for useful discussions and comments. 
The author (N.O) would like to thank Siew-Phang Ng 
 for his useful advice on our numerical analysis.

%%%%%%%%%%%%
%\newpage
%%%%%%%%%%%%%%%%%%%%%%%%%%%%%%%%

%%%%%%%%%%%%%%%
\newpage
\begin{figure}
\begin{center}
\epsfig{file=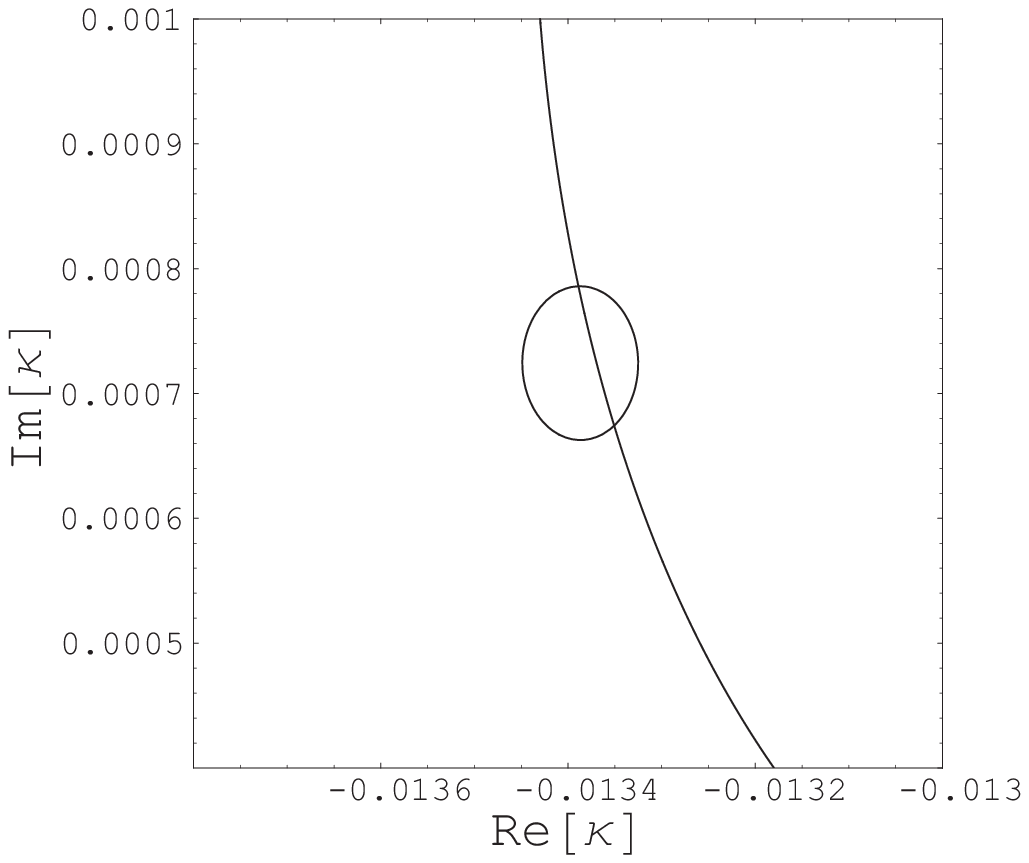, width=12cm}
\caption{
Contour plot on complex $\kappa$-plane.
The vertical line and the circle correspond to the solutions 
 of Eqs.~(\ref{cond1})  and (\ref{cond2}), respectively. 
}
\end{center}
\end{figure}
%%%%%%%%%%%%%%%%%%%%%%%%%
\newpage
\begin{figure}
\begin{center}
\epsfig{file=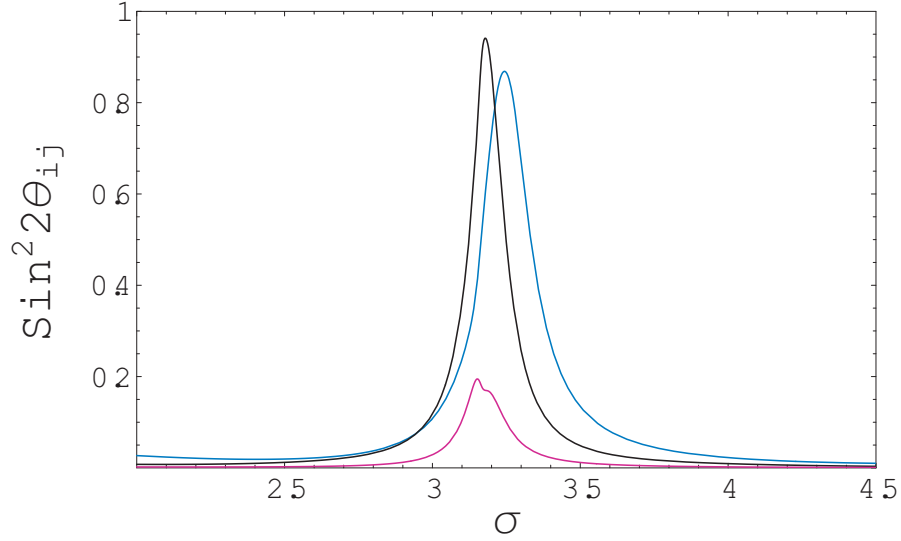, width=12cm}
\caption{
Three mixing angles in the PMNS matrix 
 as functions of $\sigma \mbox{[rad]}$. 
The graphs with the highest, middle and lowest peaks
 correspond to 
 $\sin^2 2 \theta_{2 3}$, 
 $\sin^2 2 \theta_{1 2}$ and $\sin^2 2 \theta_{1 3}$, 
 respectively. 
The plots of $\sin^2 2 \theta_{23}$ and 
 $\sin^2 2 \theta_{13}$ have the sharp peaks 
 at $\sigma \sim 3.2 [\mbox{rad}]$, 
 while $\sin^2 2 \theta_{12}$ has the sharp peak 
 at $\sigma \sim 3.3 [\mbox{rad}]$. 
}
\end{center}
\end{figure}
%%%%%%%%%%%%%%%%%%%%%%%%%
\newpage
\begin{figure}
\begin{center}
\epsfig{file=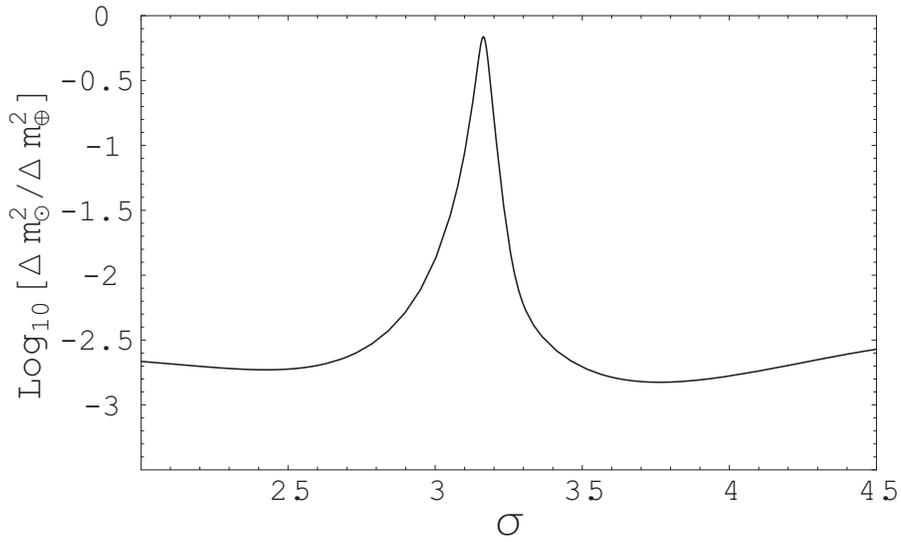, width=12cm}
\caption{
The ratio ($\mbox{Log}_{10}[\Delta m_\odot^2/\Delta m_\oplus^2]$)
 as the function of $\sigma \mbox{[rad]}$. 
}
\end{center}
\end{figure}
%%%%%%%%%%%%%%%%%%%%%%%%%
%
\end{document}